\documentclass[12pt]{article}
\usepackage{amssymb}

\begin{document}

\author{Vladimir K. Petrov\thanks{
E-mail address: petrov@earthling.net}}
\title{Analytical study of fermion determinant and chiral condensate behavior at
finite temperatures in toy model approximation.}
\date{\textit{N. N. Bogolyubov Institute for Theoretical Physics}\\
\textit{\ National Academy of Sciences of Ukraine}\\
\textit{\ 252143 Kiev, Ukraine. 4.15.1998}}
\maketitle

\begin{abstract}
Fermion determinant is computed analytically on extremely large lattices $%
N_\tau \rightarrow \infty $ in the toy model approximation in which action
is truncated so that in the Hamiltonian limit of $a_\tau \rightarrow 0$ all
terms of order $a_\tau /a_\sigma $ are discarded$.$ Chiral condensate is
studied in the area of small ( $m<<T$) quark masses.

\endabstract
\pagebreak[4]
\end{abstract}

\section{ Introduction}

Finite-temperature studies of the fermion contribution in LGT have been on
the agenda over a number of years and both experiments and analytical models
were tried to challenge the problem. A crucial issue for such studies is the
fate of chiral symmetry in continuum limit. Urgent need to perform
calculations in full QCD motivated substantial efforts made to overcome
technical problems caused by full dynamical treatment of fermions. Recent
developments in the technology of lattice gauge theory made the computation
of fermion contribution feasible. Calculations with dynamical fermions and
small lattice spacing, however, are still almost unthinkably expensive.
Therefore even crude models that allow an analytical solution might be
useful.

Unfortunately we can hardly attack the full problem, hence the
approximations which hopefully capture some of the essential features of the
physics may be considered. Here we attempt to study such a problem in
'electric toy' model approximation \cite{P}, where all terms proportional to 
$1/\xi $ has been removed \footnote{%
The anisotropy parameter $\xi $ is defined as $\xi \simeq a_\sigma /a_\tau ,$
where $a_\tau $ and $a_\sigma $ are temporal and spatial lattice spacings,
respectively.}. Several of the tools we employ exist in some form in the
earlier literature on the subject. For example in pure gluodynamics this
approximation permits to discard the magnetic part of the action \cite
{APPZ,BCAP}. In this aspect such approximation is similar to strong coupling
approximation. Moreover, effective action formally coincides with obtained
by \cite{GK,O} in $g^2>>1$ limit. On highly anisotropic lattice $\xi >>1$,
however, the condition $g^2>>1$ is superfluous and for $N_\tau >>1$ weak
coupling region is penetrable which allows to study continuum limit in this
model.

In our previous article \cite{P} pure QCD was considered in such
approximation. On extremely large lattices ( $N_\tau \rightarrow \infty $ )
the model can be solved analytically and allows to study running coupling
behavior in the continuum limit. Results obtained for the Callan-Symanzik
beta\ function are not too surprising, but hardly close to reality since
they predict trivial asymptotic freedom $g^2\sim a_\tau $. The simplest way
to move toward a more realistic case of interacting quarks and gluons is to
study dynamical fermions in the same pattern. Although it scarcely makes the
model entirely realistic , it seems not incurious to study the chiral
condensate behavior at finite temperatures within mentioned approximation.

In this paper we focus our attention on lattice formulation of QCD using
Wilson fermions, only for convenience reasons. As it is known both Wilson
and Kogut-Susskind formulations have their advantages (comprehensive
analysis of pros and contras is given in \cite{U,I}) and, of course,the
results should agree in continuum limit. Due to well known ''no-go'' theorem 
\cite{N-N} there is no straightforward way to remove fermion spectrum
degeneracy without breaking chiral invariance. For Wilson fermions chiral
symmetry is broken explicitly by the Wilson term needed to remove the
unwanted doublers on the lattice. This makes it difficult to study the
spontaneous breaking of chiral symmetry using Wilson fermions. Staggered
fermions, indeed, provide massless fermions on the lattice but the continuum
notion of chirality is not well defined for staggered fermions due to their
inherent lattice construction \cite{KBHN}.

For the fermionic part of action we choose the form suggested in \cite{HK}
(see also \cite{BRWS}) 
\begin{equation}
-S\!\!\!_F=n_f\left( a^3\sum_x\overline{\!\!\!\psi }_{x^{\prime
}}D_{x^{\prime }x}^0\psi _x+\frac{a^3}\xi \sum_{n=1}^d\sum_x\overline{%
\!\!\!\psi }_{x^{\prime }}D_{x^{\prime }x}^n\psi _x\right)  \label{sf}
\end{equation}
where $n_f$\ is the number of flavors\footnote{%
Later (unless expressly specified otherwise) we shall consider $n_f=1,$
because the generalization is evident.} and 
\begin{equation}
D_{x^{\prime }x}^n=\frac{r-\gamma _n}2\!\!\!U_n\left( x\right) \delta
_{x,x^{\prime }-n}+\frac{r+\gamma _n}2U_n^{\dagger }\left( x^{\prime
}\right) \delta _{x,x^{\prime }+n}-r\delta _{x^{\prime },x}  \label{Ds}
\end{equation}
with 
\begin{equation}
D_{x^{\prime }x}^0=\frac{r-\gamma _0}2\!\!\!U_0\left( x\right) \delta
_{x,x^{\prime }-0}+\frac{r+\gamma _0}2U_0^{\dagger }\left( x^{\prime
}\right) \delta _{x,x^{\prime }+0}-\left( ma_\tau +r\right) \delta
_{x^{\prime },x}  \label{D}
\end{equation}
Extra term with finite Wilson parameter $r$ ($0<r\leq 1$ to guarantee
positivity) becomes an irrelevant operator for ordinary fermions in the
infrared limit, while being a relevant operator for mirror fermions in the
high-energy regime, in fact , generating an effective mass for mirror
fermions\footnote{%
As it is shown below, in a given approximation it is equal to $%
m_{eff}=m+\frac 2{a_\tau }\mathop{\rm arctanh}r$ .}.

Inasmuch as in suggested approximation we omit the terms of $1/\xi $ order,
the 'toy' action will be simply $a^3\sum_x\overline{\!\!\!\psi }%
_xD_{xx^{\prime }}^0\psi _{x^{\prime }}$. Since 'toy' fermions propagation
in space direction is suppressed by factor $1/\xi $, they may hardly claim
to be dynamic in the proper sense of the word, nevertheless they bring a
nontrivial term into the effective action, which we try to calculate.

Fermion fields $\psi _x$\ can be presented as a linear combination of
creation $\mathbf{a}_{p,\sigma }^{\dagger }$\ (annihilation $\mathbf{a}%
_{p,\sigma }$) operators of the particles and antiparticles $\mathbf{b}%
_{-p,-\sigma }^{\dagger }$\ ($\mathbf{b}_{-p,-\sigma }$) with four\ momenta $%
p$\ and projection of spin $\sigma $%
\begin{eqnarray}
\psi _x &=&\sum_{p,\sigma }\left( u_{p,\sigma }\mathbf{a}_{p,\sigma
}+u_{-p,-\sigma }\mathbf{b}_{-p,-\sigma }^{\dagger }\right) ;  \nonumber \\
\overline{\!\!\!\psi }_x &=&\sum_{p,\sigma }\left( \bar u_{p,\sigma }\mathbf{%
a}_{p,\sigma }^{\dagger }+\bar u_{-p,-\sigma }\mathbf{b}_{-p,-\sigma
}\right) .
\end{eqnarray}
In a standard quantization procedure ( see, e.g. \cite{FS}) such operators
serve as integration variables, however, their combinations may be used as
well, on condition that corresponding Jacobian does not turn to zero. To
choose a convenient combinations we remind that in a standard representation 
\begin{equation}
\gamma _0=\left( 
\begin{array}{cc}
1 & 0 \\ 
0 & -1
\end{array}
\right) ;\quad \gamma _n=\left( 
\begin{array}{cc}
0 & \sigma _n \\ 
-\sigma _n & 0
\end{array}
\right) ;\quad \gamma _5=\left( 
\begin{array}{cc}
0 & -1 \\ 
-1 & 0
\end{array}
\right)
\end{equation}
projectors 
\begin{equation}
\frac{1+\gamma _0}2=\left( 
\begin{array}{cc}
1 & 0 \\ 
0 & 0
\end{array}
\right) ;\quad \frac{1-\gamma _0}2=\left( 
\begin{array}{cc}
0 & 0 \\ 
0 & 1
\end{array}
\right)
\end{equation}
divide bispinors $\psi $ into two components $\psi ^{\left( \pm \right) }$,
each including only one two-component spinor 
\begin{equation}
\psi ^{\left( +\right) }=\frac{1+\gamma _0}2\psi =\left( 
\begin{array}{c}
\psi ^{\left( +\right) } \\ 
0
\end{array}
\right) ;\quad \psi ^{\left( -\right) }=\frac{1-\gamma _0}2\psi =\left( 
\begin{array}{c}
0 \\ 
\psi ^{\left( -\right) }
\end{array}
\right)
\end{equation}
(similar to free particles and antiparticles at rest) and are the eigenvalue
of inversion$\ $ operator 
\begin{equation}
P\psi ^{\left( \pm \right) }=\pm i\psi ^{\left( \pm \right) }
\end{equation}

Taking into account that $\left( \delta _{x-\hat \mu ,x^{\prime }}\right)
^{\dagger }=\delta _{x,x-\hat \mu ^{\prime }}=\delta _{x+\hat \mu ,x^{\prime
}},$\ and presenting 
\begin{equation}
\frac{r\pm \gamma _0}2=\frac{1+\gamma _0}2\frac{r\pm 1}2+\frac{r\mp 1}2\frac{%
1-\gamma _0}2
\end{equation}
we may rewrite $\left( \ref{sf}\right) $\ in $\psi ^{\left( \pm \right) }$\
terms as 
\begin{equation}
-S_F=\sqrt{1-r^2}\left( \bar \psi _{x^{\prime }}^{\left( +\right) }\Delta
_{x^{\prime }x}^{\dagger }\psi _x^{\left( +\right) }+\bar \psi _{x^{\prime
}}^{\left( -\right) }\Delta _{x^{\prime }x}\psi _x^{\left( -\right) }\right)
;  \label{SF+}
\end{equation}
where 
\begin{equation}
\Delta _{xx^{\prime }}=\delta _{\mathbf{xx}^{\prime }}\left( \frac{e^{a_\tau
m_r}U_0\left( \mathbf{x},t\right) }2\delta _{t^{\prime },t-1}-\frac{%
e^{-a_\tau m_r}U_0^{\dagger }\left( \mathbf{x},t^{\prime }\right) }2\delta
_{t^{\prime },t+1}-m^{\prime }\delta _{t^{\prime },t}\right)  \label{del}
\end{equation}
with 
\begin{equation}
a_\tau m_r=\mathop{\rm arctanh}r  \label{mr}
\end{equation}
and 
\begin{equation}
a_\tau m^{\prime }=\frac{ma_\tau +r}{\sqrt{1-r^2}}=ma_\tau \cosh \left(
a_\tau m_r\right) +\sinh \left( a_\tau m_r\right) =\left( m+m_r\right)
a_\tau +O\left( a_\tau ^3\right)  \label{m'}
\end{equation}

By gauge transformation all $U_0\left( \mathbf{x},t\right) _{\alpha \nu }$
matrices may be diagonalized simultaneously: $U_0\left( \mathbf{x},t\right)
_{\alpha \nu }=\delta _{\alpha \nu }U_0\left( \mathbf{x},t\right) _{\alpha
\alpha },$ therefore $\Delta _{xx^{\prime }}$ in $\left( \ref{del}\right) $
is simply a set of $N$ matrices $N_\tau \times N_\tau .$ If we fix the
static gauge $U_0\left( \mathbf{x},t\right) _{\alpha \alpha }=\omega \left( 
\mathbf{x}\right) _\alpha ,$ all such matrices will depend only on $%
t-t^{\prime }$ therefore they may be diagonalized by the discrete Fourier
transformation $\exp \left( i\frac{2\pi k}{N_\tau }t\right) $ with integer $%
k $ in for periodic border conditions on fermion fields at time direction,
and half-integer $k$ in for antiperiodic ones, demanded by
Osterwalder-Schrader \cite{Z} positivity condition\footnote{%
Osterwalder and Schrader \cite{OS} developed a mathematical procedure that
allows the reconstruction of Hamiltonian and physical Hilbert space from
continuum field theory defined in Euclidean space. For Wilson formulation of
lattice gauge theory this condition was shown to hold by Osterwalder and
Seiler\cite{Z}.}. Formally a solution may be obtain with the help of the
simple equation 
\begin{equation}
\frac 1{N_\tau }\sum_k\left( \omega e^{-i\frac{2\pi k}{N_\tau }}-\frac
1\omega \exp ^{i\frac{2\pi k}{N_\tau }}\right) ^ne^{-i\frac{2\pi k}{N_\tau }%
t}=\sum_{l=-\infty }^\infty \left( 
\begin{array}{c}
n \\ 
\frac{n-t+lN_\tau }2
\end{array}
\right) \left( -1\right) ^{lB}\omega ^{-t+lN_\tau }  \label{add}
\end{equation}
where $B=0$ for periodic border conditions and $B=1$ for aperiodic ones. The
summing over $l$ is the tricky procedure and regularization is evidently
needed. If, however, we simply change $\exp \left( -i\frac{2\pi k}{N_\tau }%
\right) \rightarrow \exp \left( -i\phi \right) $ and $\frac 1{N_\tau
}\sum_k\rightarrow \int_0^{2\pi }\frac{d\phi }{2\pi }$ , it will mean $l=0$
in $\left( \ref{add}\right) $ so that the difference between periodic and
antiperiodic border conditions is lost, which is regarded as very essential
in this case. Hopefully, the computation of determinant 
\begin{equation}
\Delta =\left( 
\begin{array}{ccccc}
m^{\prime }a_\tau & \lambda U_0^{\dagger }\left( \mathbf{x},0\right) & 0 & 
... & -\lambda \left( -1\right) ^BU_0\left( \mathbf{x},N_\tau -1\right) \\ 
-\lambda U_0\left( \mathbf{x},0\right) & m^{\prime }a_\tau & \lambda
U_0^{\dagger }\left( \mathbf{x},1\right) & ... & 0 \\ 
0 & -\lambda U_0\left( \mathbf{x},1\right) & m^{\prime }a_\tau & ... & 0 \\ 
... & ... & 0 & ... & ... \\ 
0 & 0 & 0 & ... & \lambda U_0^{\dagger }\left( \mathbf{x},N_\tau -2\right)
\\ 
\lambda \left( -1\right) ^BU_0^{\dagger }\left( \mathbf{x},N_\tau -1\right)
& 0 & 0 & ... & m^{\prime }a_\tau
\end{array}
\right)
\end{equation}
where $\lambda \equiv \frac{e^{-m_ra_\tau }}2,$ in the considered extremely
simple case can be done straightforwardly and, incidentally, this does not
even need the gauge fixing (see Appendix). After the integration over $\psi
_x^{\left( \pm \right) }$ fields we get 
\begin{equation}
-S_F^{eff}=\ln \det \Delta ^{\dagger }\Delta =\sum_\alpha \ln \det \Delta
_\alpha ^{\dagger }\Delta _\alpha =\sum_\alpha \Xi _\alpha ^{*}\Xi _\alpha
\end{equation}
with 
\begin{equation}
\Xi _\alpha \left( m^{\prime }\right) =M_{N_\tau }+\left( -1\right) ^{N_\tau
}\frac 12e^{\frac{m_r}T}\Omega _\alpha +\frac 12e^{-\frac{m_r}T}\Omega
_\alpha ^{*}
\end{equation}
where 
\begin{equation}
M_{N_\tau }=\left\{ 
\begin{array}{cc}
\cosh \left( N_\tau \mathop{\rm arcsinh}m^{\prime }a_\tau \right) \simeq
\cosh \left( \frac{m^{\prime }}T\right) +O\left( a_\tau ^2\right) ; & N_\tau
=2k; \\ 
\sinh \left( N_\tau \mathop{\rm arcsinh}m^{\prime }a_\tau \right) \simeq
\sinh \left( \frac{m^{\prime }}T\right) +O\left( a_\tau ^2\right) ; & N_\tau
=2k+1.
\end{array}
\right.  \label{mn}
\end{equation}
Therefore for \textit{even }$N_\tau $we may write 
\begin{equation}
-S_F^{eff}=\sum_{\alpha =1}^N\ln \left( \rho ^2+\cos ^2\varphi _\alpha
+\left( -1\right) ^B\left( \cosh \left( \frac{m+2m_r}T\right) +\cosh \left(
\frac mT\right) \right) \cos \varphi _\alpha \right) ,  \label{E}
\end{equation}
and , respectively, for \textit{odd} $N_\tau $%
\begin{equation}
-S_F^{eff}=\sum_{\alpha =1}^N\ln \left( \rho ^2-\cos ^2\varphi _\alpha
+\left( -1\right) ^B\left( \cosh \left( \frac mT\right) -\cosh \left( \frac{%
m+2m_r}T\right) \right) \cos \varphi _\alpha \right)  \label{O}
\end{equation}
where 
\begin{equation}
\rho ^2=1+\sinh ^2\frac{m+m_r}T+\sinh ^2\frac{m_r}T.  \label{M}
\end{equation}

Both $\left( \ref{E}\right) $ and $\left( \ref{O}\right) $ can be nicely
expressed as 
\begin{equation}
-S_F^{eff}=\ln \prod_{\alpha =1}^N\left( z_r-\cos \varphi _\alpha \right)
\left( z-\left( -1\right) ^{N_\tau }\cos \varphi _\alpha \right) ;
\label{Seff}
\end{equation}
with\footnote{%
The parameter $r\ $was introduced to shift the mass of mirror fermion. As it
follows from definitions $\left( \ref{mr}\right) $ and $\left( \ref{m'}%
\right) $ $m^{\prime }-m_r=m$ and $m^{\prime }$ $+m_r=m+2$ $m_r$, so the
suggested parametrization $r=\tanh a_\tau m_r$, makes the machinery of such
shift more transparent. By increasing $m_r$ one can make the doublers as
heavy as the cutoff. It is easy to check that if $\mathrm{Im}r=0$ it does
not introduce any imaginary part into the action$.$} 
\begin{equation}
z=\left( -1\right) ^B\cosh \frac mT;\quad z_r=\left( -1\right) ^B\cosh \frac{%
m+2m_r}T
\end{equation}

In case of $SU(2)$ gauge group $\varphi _1=-\varphi _2=\frac \varphi 2=\frac
12\arccos \frac \chi 2,$ where $\chi $ is the character of fundamental
representation, so $\left( \ref{Seff}\right) $ can be easily expressed in
invariant form 
\begin{equation}
-S_F^{eff}=\ln \prod_{\alpha =1}^N\left( \cosh \frac mT+\left( -1\right)
^{B+N_\tau }\frac \chi 2\right) \left( \cosh \frac{m+2m_r}T+\left( -1\right)
^B\frac \chi 2\right) ;  \label{S2}
\end{equation}
In case of $SU(3)$ gauge group we can express $S_F^{eff}$ through the
characters of fundamental representation $\chi =\sum_{\alpha
=1}^3e^{i\varphi _\alpha }$ with the help of simple relation 
\begin{eqnarray}
\ \ \prod_{\alpha =1}^3\left( 1+pe^{i\varphi _\alpha }+qe^{-i\varphi _\alpha
}\right) &=&1+p^3+q^3-3pq+\left( p+q^2-2p^2q\right) \chi +  \nonumber \\
&&\ \ \ \left( q+p^2-2\allowbreak pq^2\right) \chi ^{*}+\ pq^2\chi
^2+p^2q\chi ^{*2}+pq\chi \chi ^{*}
\end{eqnarray}
that, e.g. in case of even $N_\tau ,$ leads to 
\begin{equation}
-S_F^{eff}=\ln f\left( \cosh \frac mT\right) +\ln f\left( \cosh \frac{m+2m_r}%
T\right) ;  \label{S3}
\end{equation}
with 
\begin{equation}
f\left( x\right) =x^3PQ^2+\frac{\chi +\chi ^{*}}2x^2PQ+\left( -1\right) ^B%
\frac{\chi ^2+(\chi ^{*})^2}8+\frac{\chi \chi ^{*}}4x;  \label{v}
\end{equation}
and 
\begin{equation}
P\left( x\right) =1+\frac{\left( -1\right) ^B}x;\quad Q\left( x\right)
=\left( -1\right) ^B-\frac 1{2x}
\end{equation}

As it can be seen from $\left( \ref{S3}\right) $ the effective action $%
S_F^{eff}$ looses $Z\left( N\right) $ invariance, already in zero order in $%
1/\xi $. Quark effects lift $Z\left( N\right) $ degeneracy and only the
phase in which Polyakov loop is real is stable. Other phases are metastable.
As it can be seen from $\left( \ref{S3}\right) $, the strength of $Z\left(
N\right) $\ symmetry breaking decreases with increasing mass 
\begin{equation}
-S_F^{eff}\simeq 2N\left( \frac{m+m_r}T-\ln 2\right) +\left( -1\right)
^B\left( \left( -1\right) ^{N_\tau }e^{-\frac mT}+e^{-\frac{m+2m_r}T}\right) 
\frac{\mathrm{Re}\chi }2;
\end{equation}
in agreement with the reasons given in \cite{MO}. This will modify the
thermodynamic properties of metastable phases, the lifetime of which
increases with decreasing strength of \ $Z\left( N\right) -$symmetry
breaking.

It is nameworthy to note that, if the integrand of partition function is
highly peaked in proximity to $\chi _{\min }$ $=\sum_\alpha \cos \varphi
_\alpha ^{\min }$, with fixed $\varphi _\alpha ^{\min }$, the partition
function has zeroes in complex $m$-plane at 
\begin{equation}
m_\alpha =\pm i\left( \varphi _\alpha +\pi n\right) T_r;\quad m_\alpha
^{\prime }=\pm i\left( \varphi _\alpha +\pi n\right) T+2m_r;\quad n=%
\mathop{\rm mod}_2\left( B+N_\tau +1\right)  \label{z}
\end{equation}

As it can be seen from $\left( \ref{mn}\right) ,$ $M_{N_\tau }$ functions
are polynomials\footnote{$M_{N_\tau }$functions are very similar to
Chebyshev polynomial, e.g., $\cosh \left( n\mathop{\rm arcsinh}\sqrt{z^2+1}%
\right) =$ $T_n\left( x\right) =\cos \left( n\arccos x\right) $} in $%
m^{\prime }a_\tau $ of order $N_\tau $, therefore, at finite $a_\tau $ the
equation $\Xi _\alpha \left( m^{\prime }\right) =0$ has $N_\tau $ solutions $%
\left\{ m_t^{\prime }\right\} $ which are, generally speaking, different and
formally can be written as: 
\begin{equation}
\frac{m^{\prime }}T=\left\{ 
\begin{array}{cc}
N_\tau \sinh \left( \frac 1{N_\tau }\mathop{\rm arccosh}\left( \frac{\left(
-1\right) ^{B+1}}2\left( e^{\frac{m_r}T}\Omega _\alpha +e^{-\frac{m_r}%
T}\frac 1{\Omega _\alpha }\right) \right) \right) ; & N_\tau =2k; \\ 
N_\tau \sinh \left( \frac 1{N_\tau }\mathop{\rm arcsinh}\frac{\left(
-1\right) ^{B+1}}2\left( -e^{\frac{m_r}T}\Omega _\alpha +e^{-\frac{m_r}%
T}\frac 1{\Omega _\alpha }\right) \right) ; & N_\tau =2k+1.
\end{array}
\right.
\end{equation}

For finite $a_\tau $ and $N_\tau $ the zeroes of $\det \Delta ^{\dagger
}\Delta $ dispersed in $m$-plane, but with $a_\tau \rightarrow 0,$and $%
N_\tau \rightarrow \infty $ they steadily concentrate around $m_\alpha $ and 
$m_\alpha ^{\prime }.$

From $\left( \ref{S2}\right) $it is evident that fermion loop contribution
increases in the parameter area where such zeroes approach the origin of
coordinates in $m$-plane (while $T$ is fixed) and simultaneously $\left| 
\frac{\chi _{\min }}2\right| ^2\rightarrow 1.$

As it is known (see e.g., \cite{BBK} and references therein), the chiral
condensate can be expressed in terms of its zeros $m_t=\frac{m_t^{\prime }}{%
\cosh \left( m_ra_\tau \right) }-\frac{\tanh \left( m_ra_\tau \right) }{%
a_\tau }$ as: 
\begin{equation}
\left\langle \bar \psi \psi \right\rangle =\frac T{V_\sigma }\sum_t\frac 1{m{%
-}m_t}.  \label{zer}
\end{equation}
where $V_\sigma $ is spatial lattice volume. The behavior of chiral
condensate in QCD can be investigated \cite{BBK} having studied the
distribution of partition function zeros in the complex quark mass plane.
Zeroes at $\left( \ref{z}\right) $, certainly, don't exhaust the whole set
of $\left\{ m_t\right\} ,$ but seem likely to represent a very important
part of it.

To finish with this issue we want to note that in a toy model approximation
the fermion contribution is simply a 'potential'-like term and can be
totally incorporated into the measure. For example in case of $SU(2)$ gauge
group the modified measure has the following form 
\begin{equation}
d\tilde \mu =\left( \cosh \frac mT+\left( -1\right) ^{B+N_\tau }\cos \frac
\varphi 2\right) \left( \cosh \frac{m+2m_r}T+\left( -1\right) ^B\cos \frac
\varphi 2\right) \sin ^2\frac \varphi 2d\varphi   \label{m2}
\end{equation}
For arbitrary flavor number $n_f$ it can be written as 
\begin{equation}
d\tilde \mu _{n_f}=\left( \cosh \frac mT+\left( -1\right) ^{B+N_\tau }\cos
\frac \varphi 2\right) ^{n_f}\left( \cosh \frac{m+2m_r}T+\left( -1\right)
^B\cos \frac \varphi 2\right) ^{n_f}\sin ^2\frac \varphi 2d\varphi 
\label{mn2}
\end{equation}
Therefore, in a toy model approximation the fermion contribution does not
seriously complicate the partition function $Z$ and thereby the computation
of any average values . Here we concentrate on the evaluation of $%
\left\langle \bar \psi \psi \right\rangle =\frac T{V_\sigma }\frac{\partial
\ln Z}{\partial m}.$

\section{Chiral condensate}

In case of $SU(3)$ gauge group in the area $\frac mT>>1$ one may write 
\begin{eqnarray}
\left\langle \bar \psi \psi \right\rangle &\simeq &\frac 1z\frac{\partial z}{%
\partial m}+\frac 1{z_r}\frac{\partial z_r}{\partial m}+\left( \frac{\left(
-1\right) ^{N_\tau }}{z^2}\frac{\partial z}{\partial m}+\frac 1{z_r^2}\frac{%
\partial z_r}{\partial m}\right) \left\langle \chi \right\rangle +... \\
\ &\simeq &\ \left( \tanh \frac mT+\tanh \frac{m+2m_r}T\right) \left(
1+\left( -1\right) ^B\left( \frac{\left( -1\right) ^{N_\tau }}{\cosh \frac mT%
}+\frac 1{\cosh \frac{m+2m_r}T}\right) \left\langle \chi \right\rangle
\right)  \nonumber
\end{eqnarray}
which means strong breakdown of chiral symmetry. Therefore, gluon
environment influence on $\left\langle \bar \psi \psi \right\rangle $ value
became inappreciable when $\frac mT$ infinitely increases.

To compute $\left\langle \bar \psi \psi \right\rangle $ in the area $\frac
mT<<1;\frac{m_r}T<<1$, we use 'mean spin method' \cite{ABPZ} which in fact
is a modification of the method of quasiaverages \cite{NNB} 
\begin{equation}
\prec A\succ _{\bar \chi }=\int A\delta \left( \frac{\bar \chi }2N_\sigma
^d-\sum_{\mathbf{x}}^{N_\sigma ^d}\cos \varphi _{\mathbf{x}}\right) \prod_{%
\mathbf{x}}d\tilde \mu _{\mathbf{x}}
\end{equation}
The effective action $\bar S=\bar S\left( \bar \chi \right) $ can be
computed as 
\begin{equation}
e^{-\bar S\left( \bar \chi \right) }\equiv \frac{\prec e^{-S}\succ _{\bar
\chi }}{\prec 1\succ _{\bar \chi }}=\sum_{n=0}^\infty \frac{\left( -1\right)
^n}{n!}\frac{\prec S^n\succ _{\bar \chi }}{\prec 1\succ _{\bar \chi }}
\end{equation}
and in the lowest order can be obtained simply by substitution $2\cos
\varphi _{\mathbf{x}}\rightarrow \bar \chi .$ In such approximation the
results coincide with a mean field method. Ordinary averages may be
evaluated by integration over $\bar \chi .$

Let us consider for simplicity the case of $SU(2)$ gauge group were the
gluonic part of action for a 'toy' model \cite{P} can be written as\footnote{%
As we have already mentioned above it coincides with the action suggested in 
\cite{GK,O} (see also \cite{BCAP}).} 
\begin{equation}
-S=4\tilde \beta \sum_{\mathbf{x,n}}\cos \frac{\varphi _{\mathbf{x}}}2\cos 
\frac{\varphi _{\mathbf{x+n}}}2=\tilde \beta \sum_{\mathbf{x,n}}\chi _{%
\mathbf{x}}\chi _{\mathbf{x+n}}  \label{ac}
\end{equation}
and fermion contribution is incorporated in the measure $d\tilde \mu $ given
by $\left( \ref{m2}\right) .$

Since we introduced the collective variable $\bar \chi =\frac 2{N_\sigma
^d}\sum_{\mathbf{x}}\cos \varphi _{\mathbf{x}}$ the corresponding Jacobian $%
\prec 1\succ _{\bar \chi }$ must be computed. Having written 
\begin{equation}
\delta \left( \frac{\bar \chi }2N_\sigma ^d-\sum_{\mathbf{x}}\cos \varphi _{%
\mathbf{x}}\right) =\int_{c-i\infty }^{c+i\infty }\exp \left\{ s\left( \frac{%
\bar \chi }2N_\sigma ^d-\sum_{\mathbf{x}}\cos \varphi _{\mathbf{x}}\right)
\right\} \frac{ds}{2\pi i}
\end{equation}
we find 
\begin{equation}
\prec 1\succ _{\bar \chi }=\exp \left( -N_\sigma ^d\frak{L}\right)
=\int_{c-i\infty }^{c+i\infty }\exp \left\{ N_\sigma ^d\left( \frac{\bar
\chi }2s-L\left( s\right) \right) \right\} \frac{ds}{2\pi i}
\end{equation}
where\footnote{%
Here we take $\left( B=0;N_\tau =2k\right) $ as an example, other cases can
be easily obtained after the substitution of $z\rightarrow -z$ or $%
z_r\rightarrow -z_r.$} 
\begin{equation}
e^{-L\left( s\right) }=\int_0^{4\pi }\exp \left\{ s\cos \frac \varphi
2\right\} d\tilde \mu =\left( z_r-\frac \partial {\partial s}\right) \left(
z-\frac \partial {\partial s}\right) \frac{I_1\left( s\right) }s
\end{equation}
or 
\begin{equation}
-L\left( s\right) =\ln \left( \left( z_rz+1+\frac 2s\left( z_r+z\right)
+\frac 6{s^2}\right) \frac{I_1\left( s\right) }s-\left( z_r+z+\frac
3s\right) \frac{I_0\left( s\right) }s\right)
\end{equation}
Due to the fact that $N_\sigma ^d\rightarrow \infty ,$ the main contribution
into $\frak{L}$ comes from $s\sim 0,$ so 
\begin{equation}
-L\left( s\right) \simeq \ln \left( \frac{z_rz+\frac 14}2\right) -\frac{%
3\eta s}4+\frac{s^2}{16\beta _0}+O\left( s^3\right)
\end{equation}
with 
\begin{equation}
\beta _0=\frac{\left( z_rz+\frac 14\right) ^2}{\left( z_r-\frac 12\right)
\left( z{}^2-\frac 12\right) +z_rz\left( z_rz+\frac 12\right) };
\end{equation}
and 
\begin{equation}
\eta =\frac 13\frac{z_r+z}{z_rz+\frac 14}.
\end{equation}

As far as $N_\sigma ^d\rightarrow \infty ,$ the function $\frak{L}$ may be
computed by the steepest descent method, with good accuracy. The saddle
point $s_0=s_0\left( \bar \chi \right) $ is given by the equation 
\begin{equation}
\frac{\partial L\left( s_0\right) }{\partial s_0}\simeq \frac 34\eta -\frac{%
s_0}{8\beta _0}+O\left( s_0^2\right) =\frac{\bar \chi }2
\end{equation}
so 
\begin{equation}
s_0\simeq 4\beta _0\left( \frac 32\eta -\bar \chi \right)  \label{So}
\end{equation}
and we obtain 
\begin{equation}
-\frak{L}\left( s_0\right) \simeq -L\left( s_0\right) +s_0\bar \chi \simeq
-3\beta _0\left( \bar \chi -\frac 32\eta \right) \left( \bar \chi -\frac
12\eta \right) ;
\end{equation}
Therefore we can finally write 
\begin{equation}
-\bar S\left( \bar \chi \right) =3\left( \tilde \beta -\beta _0\right) \bar
\chi ^2+6\eta \beta _0\bar \chi +const
\end{equation}

From the conditions 
\begin{equation}
-\bar S\left( \bar \chi \right) ^{\prime }=\ 6\left( \tilde \beta -\beta
_0\right) \bar \chi +6\eta \beta _0;\quad -\bar S\left( \bar \chi \right)
^{\prime \prime }=6\left( \tilde \beta -\beta _0\right) ;
\end{equation}
we see that $\bar S\left( \bar \chi \right) $ has the minimum for $\tilde
\beta <\beta _0$ at 
\begin{equation}
\ \bar \chi _{\min }=\bar \chi _0\equiv \ \frac \eta {1-\tilde \beta /\beta
_0}.  \label{s-p}
\end{equation}
This solution $\left( \ref{s-p}\right) $ dominate for $\bar \chi _0^2<4$ ,
but at points $\bar \chi _0=\pm 2$ corresponding to 
\begin{equation}
\tilde \beta =\beta _c\equiv \beta _0\left( 1-\frac 12\left| \eta \right|
\right) ;  \label{cr}
\end{equation}
the regime changes. For $\tilde \beta >\beta _c$ the value $\bar \chi _{\min
}^2$ keeps to $4$, so 
\begin{equation}
\bar \chi _{\min }^2=\min \left\{ 4;\bar \chi _0^2\right\} ;\quad sign\left(
\bar \chi _{\min }\right) =sign\left( \eta \right)  \label{m}
\end{equation}
As it is seen from $\left( \ref{cr}\right) $, 'toy' fermions contribution
leads to a shift of the effective coupling\footnote{%
It can be shown that such shift becomes a bit larger for a number of
flavours $n_f$ $=2$, but leaves $\beta _c>0,$ therefore does not wash out
the phase transition.}. For all $m_r,$ both $\beta _0$ and $\left| \eta
\right| $ increase with increasing $m$. This leads to decreasing $\beta
_c=\beta _c\left( \frac mT\right) $ at small $m$ , but for $m\gtrsim T$ $\ $%
it starts to slowly increase to the asymptotic value $\beta _c\left( \infty
\right) =\frac 12.$

An average value of $\left\langle Q\right\rangle $ of some $Q=Q\left( \cos
\frac \varphi 2\right) $ can be computed simply as 
\begin{equation}
\left\langle Q\right\rangle =\left( \frac{\int_0^{4\pi }Q\left( \cos \frac
\varphi 2\right) \exp \left\{ s\cos \frac \varphi 2\right\} d\tilde \mu }{%
\int_0^{4\pi }\exp \left\{ s\cos \frac \varphi 2\right\} d\tilde \mu }%
\right) _{s=s_0}=\left( e^{L\left( s\right) }Q\left( \frac \partial
{\partial s}\right) e^{-L\left( s\right) }\right) _{s=s_0}
\end{equation}
with 
\begin{equation}
s_0\simeq 4\beta _0\left( \frac 32\eta -\bar \chi _{\min }\right)  \label{Sm}
\end{equation}
where $\bar \chi _{\min }$ is defined by $\left( \ref{m}\right) $.

In particular, taking into account that 
\begin{equation}
\left\langle \frac 1{z_r-\cos \frac \varphi 2}\right\rangle =e^{L\left(
s\right) }\left( z-\frac \partial {\partial s}\right) \frac{I_1\left(
s\right) }s=\frac z{z_rz+\frac 14}-\beta _0\left( \bar \chi _{\min }-\frac
32\eta \right) \frac{z-\frac 14}{\left( z_rz+\frac 14\right) ^2}
\end{equation}
we obtain for $\left\langle \bar \psi \psi \right\rangle $%
\begin{equation}
\left\langle \bar \psi \psi \right\rangle =\frac{\partial z}{\partial m}%
\left\langle \frac 1{z+\left( -1\right) ^{B+N_\tau }\cos \frac \varphi
2}\right\rangle +\frac{\partial z_r}{\partial m}\left\langle \frac
1{z_r+\left( -1\right) ^B\cos \frac \varphi 2}\right\rangle  \label{con}
\end{equation}
which for the case considered $\left( \ref{con}\right) $ gives for the even $%
N_\tau $%
\begin{equation}
\left\langle \bar \psi \psi \right\rangle =\frac{24}{11}\frac{m+m_r}T\left(
1-\frac{\left| \bar \chi _{\min }\right| }3\right)  \label{e}
\end{equation}
and for odd $N_\tau $%
\begin{equation}
\left\langle \bar \psi \psi \right\rangle \simeq \frac 83\frac mT+\frac 43%
\frac{m_r}T\left( 1-\frac{\left| \chi _{\min }\right| }2\right)  \label{o}
\end{equation}
It is easy to see that at critical point (where $\left| \bar \chi _{\min
}\right| \simeq 2$) both $\left( \ref{e}\right) $ and $\left( \ref{o}\right) 
$ became comparatively small but still remain finite.

A very important feature of the chiral condensate is the behavior of $%
\langle \bar \psi \psi \rangle \ $with decreasing quark mass $m$. Excellent
fit of experimental data (in the symmetric phase of QED$_4$) in the fermion
mass range $0.01-0.06$ is given in \cite{ACGLP} by the relation 
\begin{equation}
\langle \bar \psi \psi \rangle \times \ln ^\varepsilon {\frac 1{\langle \bar
\psi \psi \rangle }}\sim m;\quad \varepsilon \sim \frac 14
\end{equation}
The above roughly agrees with $\left( \ref{e}\right) $ that predicts that $%
\langle \bar \psi \psi \rangle \sim \frac mT.$ It easy to check that the
data in \cite{B-W} (obtained for $SU(3)$ group with flavors number $n_f=2$
on lattice $N_\tau =6$ ) are consistent with $\left\langle \bar \psi \psi
\right\rangle \simeq 6am=\frac mT$ , which gives a reason to believe that $%
\left( \ref{e}\right) $ indeed provides a reasonable description of the data.

\section{Discussion}

\textbf{\ }It is commonly assumed that, quenched lattice QCD studies using
Wilson-Dirac fermions bring about large statistical errors in calculations
involving very light quarks. Indeed, in quenched simulations, using Wilson
fermions, lattice spacing effects were shown to be a major problem \cite{J-W}%
. Fortunately, in the parameter area available for modern computers , sea
quark contribution does not introduce substantial changes in MC data, except
on occasions. For example, comparison of spectroscopy results obtained with
dynamical and quenched fermions does not reveal any dramatic differences.
Apparently at the parameter values of the simulation the sea quarks simply
do not affect spectroscopy above the five to ten per cent level \cite{B-W}.
The measurements of valence observables in an almost-quenched system show
the physics with light dynamical quarks to be very similar to quenched
physics with the same masses \cite{KE}. The comparison between Wilson and
clover actions showed that the results were agreeable \cite{CEHS} and no
significant impact of unquenching was observed \cite{S-V} on the hyperfine
splitting as well. Therefore, the major part of MC data helps to trace the
sea quark contribution, rather than obligatory makes us take the latter into
account.

Nevertheless,when the area of small quark masses and lattice spacing will be
accessible in MC simulations, the situation presumably may change. Even with
current limited computer capacity, there are some parameter areas where the
sea quark contribution is considerable. Indeed, in \cite{H-S} effects of
dynamical fermions were singled out within the confinement phase, by
comparison of the quenched and full formulations of compact QED with Wilson
fermions. It was established, that in the strong coupling limit ($\beta =0$)
the quenched theory was a good approximation of the full one but appeared to
be in sharp contrast with it at $\beta =0.8$. At such $\beta $ the physics
changes significantly and the formation of metastable states was observed 
\cite{H-S} presumably due to the presence of dynamical fermions.

We hope that suggested toy model approximation will help not only to
estimate the sea quark contribution at lattices with size and spacings
unaccessible for MC experiment, but may help to qualify the parameter area
where the presence of dynamical fermions leads to particularly appreciable
effects.

\section{Conclusions}

Suggested approach to the problem of dynamical fermion effect estimation is
based on the discard of all terms of $1/\xi $ order in the action$.$
Therefore, only quark lines winding the lattice in the time direction are
possible. The model allows an analytical solution for small values of
lattice spacings and light quark masses on infinitely large lattices, where
MC simulations easily become prohibitively costly. For the vanishing
chemical potential $\mu ,$ the fermion determinant remains real and can be
incorporated into the measure. Although such contribution is almost trivial,
it significantly changes the phase structure of the model. It is believed
that chiral symmetry violated by Wilson term\ is restored in the continuum
limit, however, no such restoration is revealed in the model at finite $%
\frac mT$ and $\frac{m_r}T.$ \ 

In fact, two issues, which are fundamental to understand the consequences of
such approximation, have not been fully clarified: a remarkable difference
between the results obtained on lattices with odd and even $N_\tau $ (see
also \cite{BPZ}) as well as very marginal alterity between periodic and
antiperiodic border conditions.

The first direct calculation of QCD properties for small quark masses was
fulfilled long ago \cite{BGSST}. Substantial diversity between periodic and
antiperiodic border conditions was established. However, it was also pointed
out that such difference might smear out with increasing lattice volume. As
it is seen from $\left( \ref{S2}\right) $ and $\left( \ref{S3}\right) $,
alterity between periodic and antiperiodic border conditions formally does
not disappear at any large $N_\tau $ and hardly will be still unobservable
at higher orders in $\frac mT$ and $\frac{m_r}T.$ As it follows from $\left( 
\ref{Seff}\right) $ $\left( -1\right) ^B$ factor that reflects the
difference between periodic and antiperiodic conditions for even $N$ may be
absorbed into $Z\left( N\right) $-transformation. Therefore, there is still
a chance that even broken $Z\left( 2\right) $ -invariance may smear the
mentioned difference, so an additional consideration of $N=3$ is desirable,

Another problem is that for large Wilson parameter $r\rightarrow 1$ (or $%
m_r\rightarrow \infty $ ) the condensate $<\bar \psi \,\psi >$ does not
vanish at any finite temperatures. Moreover, at given approximation (in
contrast to suggestion made in \cite{H-S}) chiral symmetry broken by the
Wilson term can hardly be recovered by any fine-tuning of the bare
parameters in the continuum limit.

\section{Appendix}

Expanding $F_{n+1}$%
\begin{equation}
F_{n+1}=\left( 
\begin{array}{cccccc}
q_{n+1} & b_n & 0 & ... & 0 & c_{n+1} \\ 
c_n & q_n & b_{n-1} & ... & 0 & 0 \\ 
0 & c_{n-1} & q_{n-1} & ... & 0 & 0 \\ 
... & ... & ... & ... & ... & ... \\ 
0 & 0 & 0 & ... & q_2 & b_1 \\ 
b_{n+1} & 0 & 0 & ... & c_1 & q_1
\end{array}
\right) \quad  \label{F}
\end{equation}
in elements of the first row 
\begin{equation}
F_{n+1}=q_{n+1}f_n+b_n\frac \partial {\partial b_n}F_{n+1}+c_{n+1}\frac
\partial {\partial c_{n+1}}F_{n+1},
\end{equation}
we obtain: 
\begin{equation}
b_n\frac \partial {\partial b_n}F_{n+1}=-b_nc_nf_{n-1}-\stackrel{n+1}{%
\mathrel{\mathop{\prod }\limits_{j}}}\left( -b_j\right)
\end{equation}
and 
\begin{equation}
c_{n+1}\frac \partial {\partial c_{n+1}}F_{n+1}=-c_{n+1}b_{n+1}f_n^{\left(
2\right) }-\prod_{j=1}^{n+1}\left( -c_j\right)
\end{equation}
where 
\begin{equation}
f_n=q_nf_{n-1}-b_{n-1}c_{n-1}f_{n-2}=F_n\left( b_n=c_n=0\right)
\end{equation}
and 
\begin{equation}
f_n^{\left( 2\right) }=\frac \partial {\partial q_{n+1}}\frac \partial
{\partial q_1}F_{n+1}
\end{equation}
In a simple case of $q_j$ and the product $b_jc_j$ independent of $j$, 
\begin{equation}
q_j=q;\quad b_n=\frac \lambda {\theta _n};\quad c_n=-\lambda \theta _n
\end{equation}
we can write 
\begin{equation}
f_j^{\left( 2\right) }=f_{j-1}
\end{equation}
and 
\begin{equation}
f_{j+1}-qf_j-\lambda ^2f_{j-1}=0  \label{f}
\end{equation}
Equation $\left( \ref{f}\right) $ $\ $can be easily solved by introducing
the generating function 
\begin{equation}
f\left( z\right) =\sum_{j=n_{_0}}^\infty f_jz^{j-1}
\end{equation}
This leads to 
\begin{equation}
\sum_{j=n_{_0}}^\infty \left( \frac 1zf_{j+1}z^{j+1}-qf_jz^j-\lambda
^2zf_{j-1}z^{j-1}\right) =\left( \frac 1z-q-\lambda ^2z\right) f\left(
z\right) =0  \label{eqf}
\end{equation}
where $n_{_0}$ is an arbitrary fixed number. For example for $n_{_0}$ $=2$
or $n_{_0}$ $=3$ from $\left( \ref{F}\right) $we easily find 
\begin{equation}
\begin{array}{ccc}
f_2=q^2+\lambda ^2 &  & f_3=q^3+2\lambda ^2q
\end{array}
\end{equation}
therefore, general solution may be written as 
\begin{equation}
f_n=\frac{f_{n_{_0}}}{2\pi i}\displaystyle \oint \frac{z^{-n-1}dz}{\frac
1z-q-\lambda ^2z}=\left\{ 
\begin{array}{cc}
2\lambda ^n\cosh \left( n\mathop{\rm arcsinh}\frac q{2\lambda }\right) ; & 
n=2k; \\ 
2\lambda ^n\sinh \left( n\mathop{\rm arcsinh}\frac q{2\lambda }\right) ; & 
n=2k+1.
\end{array}
\right.
\end{equation}
Having collected everything we may finally write for $F_n$%
\begin{equation}
\frac{F_n}{2\lambda ^n}=\left\{ 
\begin{array}{cc}
\cosh \left( n\mathop{\rm arcsinh}\frac q{2\lambda }\right) -\frac{\frac
1\Theta +\Theta }2; & n=2k; \\ 
\sinh \left( n\mathop{\rm arcsinh}\frac q{2\lambda }\right) -\frac{\frac
1\Theta -\Theta }2; & n=2k+1,
\end{array}
\right.  \label{Fs}
\end{equation}
where 
\begin{equation}
\Theta =\prod_{j=1}^n\theta _j
\end{equation}

\end{document}